\documentclass[11pt]{article}
\usepackage[a4paper,margin=1in]{geometry}
\usepackage[T1]{fontenc}
\usepackage{lmodern,microtype,amsmath,amssymb,amsthm,mathtools}
\usepackage{graphicx,booktabs,array}
\usepackage[authoryear,round]{natbib}
\usepackage{xurl}
\usepackage[colorlinks=true,allcolors=blue!45!black]{hyperref}
\usepackage{xcolor}
\usepackage{enumitem}
\setlist{nosep}
\newtheorem{theorem}{Theorem}[section]
\newtheorem{proposition}[theorem]{Proposition}
\newtheorem{lemma}[theorem]{Lemma}
\newtheorem{corollary}[theorem]{Corollary}
\theoremstyle{definition}
\newtheorem{definition}[theorem]{Definition}
\newtheorem{example}[theorem]{Example}
\theoremstyle{remark}

\newcommand{\E}{\mathbb{E}}
\newcommand{\Pp}{\mathbb{P}}
\newcommand{\R}{\mathbb{R}}
\newcommand{\st}{\leq_{\mathrm{st}}}
\newcommand{\phibar}{\overline{\Phi}}
\DeclareMathOperator{\Bin}{Bin}
\DeclareMathOperator{\Cov}{Cov}
\newcommand{\limitconstant}{0.033826}
\newcommand{\firstfailure}{7}
\newcommand{\bencherrorest}{1.8e-15}

\newcommand{\mcapaccuracylog}{2.8\times10^{-14}}
\newcommand{\mcapaccuracyroot}{1.7\times10^{-16}}

\hypersetup{pdftitle={Portfolio-Size Constraints on Default Correlation: Exact Coherence Tests and Sharp Asymptotic Bounds},pdfauthor={M. Palaisti}}
\title{Portfolio-Size Constraints on Default Correlation:\\Exact Coherence Tests and Sharp Asymptotic Bounds}
\author{M. Palaisti}

\date{ }
\begin{document}
\maketitle
\begin{abstract}
An absorbing default process imposes joint restrictions on cumulative default distributions across horizons. We characterize these restrictions for finite homogeneous portfolios. For binomial mixtures whose mixing distribution functions cross at most once, stochastic ordering of the entire default count is equivalent to two inequalities: the probability of no defaults must be nonincreasing and the probability of all defaults must be nondecreasing. In the one-factor Gaussian model, these inequalities give an exact interval of admissible later correlations, obtained from two scalar equations. The intervals are nested as portfolio size increases. If the earlier and later default probabilities are $p_0<p_1$ and the earlier latent correlation is $\rho_0\in(0,1)$, both interval boundaries approach $\rho_0$ at rate $(\log n)^{-1/2}$, with common leading magnitude $\rho_0\sqrt{2(1-\rho_0)}[\Phi^{-1}(p_1)-\Phi^{-1}(p_0)]$. We also give a constructive absorbing coupling, an exact sequential repair, differential conditions for smooth horizon curves, and a finite-size certificate of incompatibility. Log-domain quadrature, bracketed root finding, and deterministic numerical experiments evaluate the intervals and verify the restrictions on the full count distribution. The framework provides a feasibility analysis of prescribed horizon laws.
\end{abstract}


\section{Introduction}
An individual default forecast gives the probability that a borrower defaults by a specified horizon. A portfolio forecast describes how many borrowers may default together. Consistency across horizons requires cumulative default counts to be nondecreasing along each realized path. This pathwise restriction implies that every upper-tail probability of the count distribution is nondecreasing in maturity.

This restriction arises naturally when portfolio risk is calibrated across maturities. A borrower model can supply a cumulative default-probability curve, while a dependence model supplies a correlation at each horizon. Together they determine a sequence of portfolio default-count distributions. The feasibility of those distributions as one cumulative default process depends on their joint stochastic ordering. In particular, increasing expected defaults does not guarantee that all portfolio loss probabilities evolve consistently.

A homogeneous one-factor Gaussian model provides a tractable setting in which to study this issue. Its single-horizon count distribution is a binomial mixture, determined by a cumulative default probability and a latent correlation parameter. A finite pool permits changes in that parameter across horizons while preserving the existence of an absorbing default process. The admissible changes depend on the number of borrowers, the movement in cumulative default probability, and the earlier correlation.

The analysis is related to several strands of credit-risk research. The Gaussian default-time construction of \citet{li2000} connects individual default curves through a common dependence structure. \citet{andersen2006} studies term-structure and conditional-survival restrictions in factor models and their implications for intertemporal portfolio losses. \citet{torresetti2007} develops consistency conditions for expected tranched loss surfaces, and \citet{jackson2011} considers temporal dependence in dynamic factor copulas. The large-pool interpretation of the factor model is central to \citet{vasicek2002} and \citet{gordy2003}.

Our formulation starts from prescribed exchangeable horizon laws and asks whether a joint absorbing process exists with those laws. It permits arbitrary couplings between horizons. The resulting finite-portfolio characterization combines a two-event criterion for single-crossing binomial mixtures with exact correlation intervals in the Gaussian model. It also yields a constructive absorbing coupling, a sequential projection for incompatible correlation curves, and a differential characterization for smooth curves. The dependence of the feasible interval on portfolio size admits the expansion
\begin{equation}\label{eq:intro-rate} \rho_n^{\pm}=\rho_0\pm\frac{\rho_0\sqrt{2(1-\rho_0)}\,[\Phi^{-1}(p_1)-\Phi^{-1}(p_0)]}{\sqrt{\log n}}+O\!\left(\frac{\log\log n}{\log n}\right). \end{equation}
The common leading constant identifies the rate at which the finite-portfolio freedom in correlation contracts. It also determines the critical size at which a fixed small correlation change becomes incompatible with absorption.

The mathematical ingredients have established precedents. \citet{strassen1965} provides the classical monotone-coupling framework; \citet{shaked2007} develops stochastic-order results; \citet{boland1989} studies crossing properties of mixture distributions; and \citet{yu2009} provides ordering criteria for exponential-family mixtures. Gaussian orthant monotonicity is related to \citet{plackett1954}, and high-dimensional orthant asymptotics are studied by \citet{rinott2001}. The present results connect these tools to a finite-pool maturity-feasibility problem, with exact scalar correlation bounds and a sharp asymptotic characterization of their endpoints.

Recent credit portfolio comparisons also emphasize dependence uncertainty. \citet{ansari2026} derives convex-order comparisons and risk bounds for Bernoulli mixture models, including Gaussian thresholds. Those results address dispersion and portfolio-risk bounds under model uncertainty. For cumulative default forecasts, first-order stochastic order characterizes maturity feasibility of the count laws. The single-crossing criterion below makes that order test computationally inexpensive for the homogeneous Gaussian family.

Network-based borrower prediction provides an additional motivation. \citet{oskarsdottir2021} and \citet{zandi2025} develop predictive models using multilayer borrower relationships. Their outputs can inform horizon-specific cumulative default probabilities, while a portfolio dependence model determines the associated count distributions. The Gaussian benchmark offers a transparent feasibility diagnostic for this latter step.

We work with exchangeable borrowers and equal loss amounts per default. All horizon laws refer to the same forecasting origin and information set. The correlation parameter is the latent Gaussian correlation. The numerical section combines constructed examples, exact finite-size calculations, and deterministic sensitivity analyses, with scripts provided for replication.

\section{Default forecasting and temporal compatibility}\label{sec:background}
\subsection{Cumulative default probabilities and portfolio distributions}
Let $\tau_i$ be the first default time of borrower $i$. Its cumulative default indicator is $D_i(t)=\mathbf{1}_{\{\tau_i\leq t\}}$, so $D_i(s)\leq D_i(t)$ whenever $s<t$. This is the sense in which default is absorbing. The indicator records default history through time $t$, including borrowers who subsequently cure or repay. A current delinquency indicator records the borrower's status at the observation date and can return to zero after a cure.

All horizon distributions in this paper are assessed at one forecasting origin, under a common information set. Thus $p_i(t)=\Pp(\tau_i\leq t)$ must be nondecreasing in $t$. Forecast updating introduces an evolving information set: a revised estimate of default probability by a fixed terminal date can rise or fall as information arrives. Our compatibility conditions describe the cumulative horizon laws associated with each fixed information set.

For a homogeneous portfolio, write $p_i(t)=p(t)$ and $K(t)=\sum_{i=1}^nD_i(t)$. Marginal probabilities determine $\E K(t)=np(t)$. The joint law of the default indicators determines the remaining features of the count distribution. With equal exposures and a common deterministic loss per default, cumulative portfolio loss is proportional to $K(t)$, so the same ordering restrictions apply to its distribution.

Absorption requires every event $\{K(s)\geq k\}$ to be contained in $\{K(t)\geq k\}$ along a common realization. Hence every count-tail probability must be nondecreasing in the horizon. This is first-order stochastic dominance, expressed through the finite set of thresholds $k=1,\ldots,n$. It orders the expectations of all increasing functions of the count and, in particular, its mean; see \citet{shaked2007}.

Expected tranche losses provide an economic interpretation. \citet{torresetti2007} use maturity and attachment-point restrictions to construct expected tranched loss surfaces from CDO quotes. For the normalized count loss $L(t)=K(t)/n$, a tranche spanning $[(k-1)/n,k/n]$ has loss
\begin{equation}\label{eq:tranche-tail} \min\!\left\{\left(L(t)-\frac{k-1}{n}\right)_+,\frac1n\right\}=\frac1n\mathbf{1}_{\{K(t)\geq k\}},\qquad k=1,\ldots,n. \end{equation}
Thus count-tail ordering is equivalent to nondecreasing expected losses for every single-default-width tranche. The two extreme events in our criterion correspond to the first and last of these tranches. The Gaussian characterization gives conditions under which their monotonicity implies ordering across the entire count grid. A practical measure of an inconsistency between two horizons is
\begin{equation}\label{eq:violation} V_n=\max_{1\leq k\leq n}\bigl[\Pp(K_0\geq k)-\Pp(K_1\geq k)\bigr]_+. \end{equation}
It is the largest decline in a count-tail probability and, for a common deterministic loss amount per default, the corresponding maximum decline in expected loss across single-default-width layers divided by that loss amount. Exact feasibility corresponds to $V_n=0$.

\subsection{The Gaussian benchmark and the coupling question}
The Gaussian copula construction links individual default curves through latent normal variables; see \citet{li2000}. In the homogeneous one-factor specification, let $V,\epsilon_1,\ldots,\epsilon_n$ be independent standard normals and set
\begin{equation}\label{eq:background-factor} \begin{aligned} X_i&=\sqrt\rho\,V+\sqrt{1-\rho}\,\epsilon_i,\qquad U_i=\Phi(X_i),\\ D_i(t)&=\mathbf{1}_{\{U_i\leq p(t)\}},\qquad 0\leq\rho<1. \end{aligned} \end{equation}
Each $U_i$ is uniform, giving the required marginal probability $p(t)$. Using the same $U_i$ at every horizon makes each default indicator nondecreasing whenever $p(t)$ is nondecreasing. Conditional on the factor, the indicators at a fixed horizon are independent, and their count is binomial. Integrating over the factor produces the binomial mixture studied below, as in the conditional-independence framework of \citet{vasicek2002}.

Allowing a separate correlation at each horizon changes the distribution of the conditional default probability. A fixed-factor construction requires monotonicity at every factor realization, as in the term-structure analysis of \citet{andersen2006}. The general coupling formulation allows a joint law across horizons with the prescribed horizon marginals. It treats the temporal connection between those marginals as a modeling choice subject to absorption. Proposition~\ref{prop:coupling} provides an explicit construction for exchangeable portfolios.

\section{Absorbing default laws for an exchangeable portfolio}
Fix ordered horizons $t_0<\cdots<t_m$ and a portfolio size $n$. At horizon $t_j$, let $\nu_j$ be an exchangeable probability law on $\{0,1\}^n$. Write $D_{ij}$ for the default indicator of borrower $i$ and $K_j=\sum_{i=1}^n D_{ij}$ for the default count.

\begin{definition}
The laws $\nu_0,\ldots,\nu_m$ are absorbing-compatible if there is a joint distribution with these horizon marginals such that $D_{i,j+1}\geq D_{ij}$ almost surely for every $i$ and $j$.
\end{definition}

An exchangeable binary law is determined by the distribution of its count: conditional on $K_j=k$, each subset of $k$ defaulted names has probability $\binom nk^{-1}$. For integer-valued random variables supported on $\{0,\ldots,n\}$, we write $K\st K'$ if $\Pp(K\geq k)\leq\Pp(K'\geq k)$ for $k=1,\ldots,n$.

\begin{proposition}\label{prop:coupling}
The laws $\nu_0,\ldots,\nu_m$ are absorbing-compatible if and only if $K_0\st\cdots\st K_m$. If this condition holds, an exchangeable absorbing coupling can be constructed using one uniform random variable and one independent uniform permutation of the borrowers.
\end{proposition}
\begin{proof}
Necessity follows by summing $D_{i,j+1}\geq D_{ij}$. For sufficiency, let $F_j$ be the count distribution function and define $K_j=F_j^{-1}(U)$ for one $U$ uniform on $(0,1)$, using the left quantile. Stochastic ordering gives $K_0\leq\cdots\leq K_m$ almost surely. Let $R_1,\ldots,R_n$ be the ranks of an independent uniform permutation, and put $D_{ij}=\mathbf{1}_{\{R_i\leq K_j\}}$. Conditional on $K_j=k$, the defaulted set is uniform among all subsets of size $k$. Thus the horizon law is $\nu_j$, and every indicator is nondecreasing in $j$.
\end{proof}

The coupling defines a default-time model on the given grid by assigning each borrower the first horizon at which its indicator becomes one. Borrowers already in default at $t_0$ can be assigned time $t_0$, and those surviving $t_m$ a time after $t_m$. A given family of horizon distributions generally admits multiple temporal laws.

\begin{example}\label{ex:arbitrary}
Let $n=3$, let $K_0=2$ surely, and let $K_1$ equal $1$ or $3$ with probabilities $2/5$ and $3/5$. Both no-default probabilities are zero, while the all-default probability rises from zero to $3/5$. The mean count rises from $2$ to $11/5$, and $\Pp(K_1\geq2)=3/5<1=\Pp(K_0\geq2)$. This interior-tail inequality violates the ordering required for absorption. Uniform defaulted subsets conditional on these counts give exchangeable binary laws with the stated properties.
\end{example}

The following section gives the additional structure under which the two extreme events do suffice.

\section{A two-event criterion for binomial mixtures}
For $a\in\{0,1\}$, suppose
\begin{equation}\label{eq:mixture} K_a\mid Q_a\sim\Bin(n,Q_a),\qquad 0<Q_a<1, \end{equation}
and denote the distribution function of $Q_a$ by $H_a$. The variable $Q_a$ represents the default probability conditional on the common environment. If $p_a=\E Q_a$, the mixture has
\begin{equation}\label{eq:mixture-moments} \E K_a=np_a,\qquad \operatorname{Var}(K_a)=np_a(1-p_a)+n(n-1)\operatorname{Var}(Q_a). \end{equation}
The second term records the contribution of common uncertainty to count dispersion. Different mixing laws can therefore preserve the mean while changing the likelihood of both mild and severe default outcomes. The criterion below compares their entire count distributions. A function has at most one sign change if it has constant sign, or changes once from nonnegative to nonpositive, or once in the reverse direction. Zeros are allowed.

\begin{theorem}\label{thm:two}
Assume that $H_0-H_1$ has at most one sign change on $(0,1)$. Then $K_0\st K_1$ if and only if
\begin{equation}\label{eq:two} \E(1-Q_1)^n\leq\E(1-Q_0)^n,\qquad \E Q_1^n\geq\E Q_0^n. \end{equation}
\end{theorem}
\begin{proof}
Necessity follows from the count thresholds $1$ and $n$. For sufficiency, put $g=H_0-H_1$ and $w_k(q)=q^{k-1}(1-q)^{n-k}$. Integration by parts gives
\begin{equation}\label{eq:tailintegral} T_k:=\Pp(K_1\geq k)-\Pp(K_0\geq k)=n\binom{n-1}{k-1}\int_0^1 w_k(q)g(q)\,dq. \end{equation}
Indeed, the derivative of the binomial tail with respect to $q$ is $n\binom{n-1}{k-1}w_k(q)$, and the boundary terms vanish. Conditions \eqref{eq:two} say $T_1,T_n\geq0$.

If $g\geq0$, every integral is nonnegative. If $g\leq0$, $T_1\geq0$ and positivity of $w_1$ on $(0,1)$ force $g=0$ almost everywhere, so every integral vanishes. Suppose next that $g$ is nonnegative below a crossing point $q_*$ and nonpositive above it. The ratio $w_n/w_k=[q/(1-q)]^{n-k}$ is nondecreasing. Consequently
\begin{equation}\label{eq:ratio1} w_n(q)g(q)\leq\frac{w_n(q_*)}{w_k(q_*)}w_k(q)g(q),\qquad 0<q<1. \end{equation}
Since the integral on the left is nonnegative, the integral of $w_kg$ is also nonnegative. If $g$ is nonpositive below $q_*$ and nonnegative above it, use instead the nondecreasing ratio $w_k/w_1=[q/(1-q)]^{k-1}$ to obtain
\begin{equation}\label{eq:ratio2} w_k(q)g(q)\geq\frac{w_k(q_*)}{w_1(q_*)}w_1(q)g(q). \end{equation}
Now $T_1\geq0$ implies the required inequality. The cases in which the crossing occurs only at a boundary reduce to constant sign. Hence $T_k\geq0$ for every $k$.
\end{proof}

The proof uses a variation-diminishing argument for the binomial kernel, in the tradition of \citet{boland1989}. The condition is stated directly for the mixing distribution functions, which are available in closed form for location-scale probit mixtures. The conclusion replaces the $n$ separate count-tail comparisons with two extreme-event probabilities.

\begin{corollary}\label{cor:location}
Let $Z$ have a continuous strictly increasing distribution function $F$ on $\R$, let $G:\R\to(0,1)$ be a continuous strictly increasing bijection, and let $Q_a=G(\mu_a+\sigma_a Z)$ with $\sigma_a>0$. The equivalence in Theorem~\ref{thm:two} holds.
\end{corollary}
\begin{proof}
The distribution function is $H_a(q)=F((G^{-1}(q)-\mu_a)/\sigma_a)$. Because $F$ is strictly increasing, the sign of $H_0(q)-H_1(q)$ is the sign of
\begin{equation}\label{eq:affine} \left(\frac1{\sigma_0}-\frac1{\sigma_1}\right)G^{-1}(q)-\frac{\mu_0}{\sigma_0}+\frac{\mu_1}{\sigma_1}. \end{equation}
This affine function of $G^{-1}(q)$ has at most one sign change.
\end{proof}

The result applies to both probit-normal and logit-normal mixing. It also applies with a nonnormal full-support factor, provided the same factor distribution and link are used at the two horizons. This common location-scale representation supplies the single-crossing property used in the theorem. More general factor families and heterogeneous borrower probabilities require corresponding conditions on the mixing distributions and count laws.

\section{Exact Gaussian correlation intervals}
Let $Z$ be standard normal and define
\begin{equation}\label{eq:Q} Q(p,\rho)=\Phi\!\left(\frac{c(p)+\sqrt\rho\,Z}{\sqrt{1-\rho}}\right),\qquad c(p)=\Phi^{-1}(p),\quad 0<p<1,\quad 0<\rho<1. \end{equation}
Then $\E Q(p,\rho)=p$. For instance, if $\epsilon$ is an independent standard normal, $\Pp(\sqrt{1-\rho}\,\epsilon-\sqrt\rho Z\leq c(p))=p$. The resulting binary vector is the homogeneous one-factor Gaussian default model introduced in Section~\ref{sec:background}. The sign of $Z$ is immaterial.

The parameter $\rho$ is the correlation of the latent normals. For distinct borrowers, the resulting correlation of their default indicators is
\begin{equation}\label{eq:indicator-correlation} \operatorname{Corr}(D_i,D_j)=\frac{\Phi_2(c(p),c(p);\rho)-p^2}{p(1-p)},\qquad i\ne j, \end{equation}
where $\Phi_2(\cdot,\cdot;\rho)$ denotes the bivariate standard normal distribution function with correlation $\rho$. Thus a constant latent correlation can coexist with changing default-indicator correlation as the cumulative probability $p$ increases. The interval restrictions below concern the latent parameter. The conditional default probability in \eqref{eq:Q} has probit location and scale
\begin{equation}\label{eq:musigma} \mu(p,\rho)=\frac{c(p)}{\sqrt{1-\rho}},\qquad \sigma(\rho)=\sqrt{\frac\rho{1-\rho}}. \end{equation}
Define the extreme-event probabilities
\begin{equation}\label{eq:AS} A_n(p,\rho)=\E Q(p,\rho)^n,\qquad S_n(p,\rho)=\E[1-Q(p,\rho)]^n=A_n(1-p,\rho). \end{equation}
At $\rho=0$ these extend continuously to $A_n(p,0)=p^n$ and $S_n(p,0)=(1-p)^n$.

\begin{lemma}\label{lem:monotonic}
For $n\geq2$, $A_n(p,\rho)$ and $S_n(p,\rho)$ are continuous and strictly increasing in $\rho\in[0,1)$. At fixed $\rho<1$, $A_n$ is strictly increasing in $p$, while $S_n$ is strictly decreasing in $p$.
\end{lemma}
\begin{proof}
The assertions in $p$ follow directly from \eqref{eq:Q}. For correlation, $A_n$ is the probability that all coordinates of an equicorrelated standard normal vector lie below $c(p)$. The identity of \citet{plackett1954} expresses the derivative with respect to any off-diagonal correlation as the bivariate normal density at the two corresponding thresholds, multiplied by the conditional probability that the remaining coordinates lie below their thresholds. Both factors are strictly positive for a nonsingular correlation matrix and finite thresholds. Along the equicorrelation path, the derivative is the sum of these positive quantities over pairs. Continuity at zero follows by dominated convergence in \eqref{eq:Q}. Apply the same argument at $1-p$ for $S_n$.
\end{proof}

Fix a correlation cap $\bar\rho<1$, an earlier pair $(p_0,\rho_0)$ with $0<\rho_0<\bar\rho$, and a later cumulative probability $p_1\geq p_0$. Let $I_n(p_0,\rho_0;p_1)$ be the set of later correlations in $[0,\bar\rho]$ for which the two Gaussian horizon laws are absorbing-compatible.

\begin{theorem}\label{thm:band}
For $n\geq2$, $I_n(p_0,\rho_0;p_1)=[\ell_n,u_n]$, where
\begin{equation}\label{eq:lower} \ell_n=\begin{cases}0,&p_1^n\geq A_n(p_0,\rho_0),\\ \text{the unique }r\in(0,\rho_0]\text{ with }A_n(p_1,r)=A_n(p_0,\rho_0),&\text{otherwise},\end{cases} \end{equation}
and
\begin{equation}\label{eq:upper} u_n=\begin{cases}\bar\rho,&S_n(p_1,\bar\rho)\leq S_n(p_0,\rho_0),\\ \text{the unique }r\in[\rho_0,\bar\rho)\text{ with }S_n(p_1,r)=S_n(p_0,\rho_0),&\text{otherwise}.\end{cases} \end{equation}
The interval contains $\rho_0$. If $p_1=p_0$, it is the singleton $\{\rho_0\}$. If $p_1>p_0$, it contains an open neighborhood of $\rho_0$. For any finite sequence of such Gaussian horizon laws, absorbing compatibility is equivalent to satisfaction of these intervals at every adjacent pair of horizons.
\end{theorem}
\begin{proof}
For positive correlations, Corollary~\ref{cor:location} gives equivalence to $A_n(p_1,r)\geq A_n(p_0,\rho_0)$ and $S_n(p_1,r)\leq S_n(p_0,\rho_0)$. Lemma~\ref{lem:monotonic} yields the stated roots and the interval between them. The fixed-correlation choice $r=\rho_0$ satisfies both inequalities because $p_1\geq p_0$. At equality of the probabilities, strict monotonicity in correlation forces $r=\rho_0$. For $p_1>p_0$, both inequalities are strict at $\rho_0$, so continuity gives a neighborhood.

At $r=0$, $Q_1=p_1$ surely. Its distribution function differs from $H_0$ by a function that is nonnegative below $p_1$ and nonpositive above it. Theorem~\ref{thm:two} therefore covers this endpoint as well. Finally, pairwise stochastic order is transitive and Proposition~\ref{prop:coupling} supplies one joint absorbing coupling on the whole grid.
\end{proof}

The root equations admit stable numerical evaluation in logarithmic form. The lower boundary enforces a nondecreasing probability of universal default. The upper boundary enforces a nonincreasing probability of universal survival. At fixed $p$, increasing dependence raises both extreme-event probabilities, giving each boundary a distinct role.

\begin{corollary}\label{cor:direction}
If $p_1\geq p_0$ and $\rho_1\geq\rho_0$, compatibility is equivalent to $S_n(p_1,\rho_1)\leq S_n(p_0,\rho_0)$. If $p_1\geq p_0$ and $\rho_1\leq\rho_0$, it is equivalent to $A_n(p_1,\rho_1)\geq A_n(p_0,\rho_0)$.
\end{corollary}
\begin{proof}
In the first case the all-default inequality holds automatically by monotonicity in both arguments. In the second case the no-default inequality holds automatically, since increasing $p$ and decreasing $\rho$ both decrease $S_n$. Apply Theorem~\ref{thm:band}.
\end{proof}

\begin{proposition}\label{prop:nesting}
The intervals are nested in portfolio size: $I_{n+1}(p_0,\rho_0;p_1)\subseteq I_n(p_0,\rho_0;p_1)$ for every $n\geq2$.
\end{proposition}
\begin{proof}
An absorbing coupling for $n+1$ borrowers remains absorbing after one borrower is deleted. Each restricted horizon law is still the Gaussian binomial-mixture law with the same $(p,\rho)$ parameters and $n$ borrowers. Hence any correlation feasible for $n+1$ is feasible for $n$.
\end{proof}

Consequently, for a fixed pair of parameter values, the set of feasible integer portfolio sizes is an initial segment, possibly all positive integers. Portfolio enlargement imposes successively tighter restrictions on the same marginal probabilities and latent correlations.

\subsection{Sequential repair}
Suppose nondecreasing cumulative probabilities $p_0,\ldots,p_m$ and raw correlations $\widehat\rho_0,\ldots,\widehat\rho_m$ in $[0,\bar\rho]$ have been supplied. Set $\rho_0^*=\widehat\rho_0$. Given $\rho_{j-1}^*$, calculate the interval $[\ell_j,u_j]$ for the transition from $(p_{j-1},\rho_{j-1}^*)$ to $p_j$, and define
\begin{equation}\label{eq:repair} \rho_j^*=\min\{u_j,\max\{\ell_j,\widehat\rho_j\}\},\qquad j=1,\ldots,m. \end{equation}
The formulas extend to an earlier correlation of zero by the same extreme-event inequalities.

\begin{proposition}\label{prop:repair}
The repaired sequence is absorbing-compatible and preserves every $p_j$. Conditional on the preceding repaired pair, \eqref{eq:repair} is the unique minimizer of $|r-\widehat\rho_j|^a$ over the feasible interval for every $a>0$.
\end{proposition}
\begin{proof}
The interval is nonempty because retaining the preceding correlation is feasible. Projection selects a point in it. Theorem~\ref{thm:band} and Proposition~\ref{prop:coupling} give compatibility. The individual probability equals $p_j$ for every correlation in \eqref{eq:Q}. The last assertion is the elementary closest-point property of a closed interval.
\end{proof}

Each projection minimizes the current adjustment conditional on the preceding repaired pair. Earlier choices determine the feasible intervals at later maturities. A global calibration would optimize those choices jointly across the full horizon grid, and selection of a temporal law would introduce further modeling criteria.

\section{Smooth horizon curves}
The finite-grid characterization has a differential counterpart. Suppose $\mu(t)$ and $\sigma(t)>0$ are continuously differentiable on a compact interval, with $\sigma$ bounded away from zero. Let $Z$ be standard normal, $Q_t=\Phi(\mu(t)+\sigma(t)Z)$, and $K_t\mid Q_t\sim\Bin(n,Q_t)$. Define
\begin{equation}\label{eq:weights} J_k(t)=\E\!\left[\phi(\mu+\sigma Z)Q_t^{k-1}(1-Q_t)^{n-k}\right],\qquad m_k(t)=\frac{\E[Z\phi(\mu+\sigma Z)Q_t^{k-1}(1-Q_t)^{n-k}]}{J_k(t)}. \end{equation}
Here and below $\mu$ and $\sigma$ inside expectations are evaluated at $t$. These quantities are finite and $J_k(t)>0$.

\begin{proposition}\label{prop:speed}
All count tails $\Pp(K_t\geq k)$ are nondecreasing in $t$ if and only if, at every $t$,
\begin{equation}\label{eq:speed} \dot\mu(t)+\dot\sigma(t)m_1(t)\geq0,\qquad \dot\mu(t)+\dot\sigma(t)m_n(t)\geq0. \end{equation}
Moreover, $m_1(t)<\cdots<m_n(t)$ for $n\geq2$.
\end{proposition}
\begin{proof}
Differentiation under the integral is justified by Gaussian integrability and boundedness of the parameters and their derivatives on the compact interval. The binomial-tail derivative gives
\begin{equation}\label{eq:deriv} \frac{d}{dt}\Pp(K_t\geq k)=n\binom{n-1}{k-1}J_k(t)[\dot\mu(t)+\dot\sigma(t)m_k(t)]. \end{equation}
Consider the probability measure with density proportional to the integrand defining $J_k$. Passing from $k$ to $k+1$ tilts it by the strictly increasing function $Q_t/(1-Q_t)$ of $Z$. Hence the mean of $Z$ strictly increases. More explicitly, for an integrable increasing tilt $h$, the change in the mean is $\Cov(Z,h(Z))/\E h(Z)>0$; the covariance is half the expectation of $(Z-Z')[h(Z)-h(Z')]$ for independent copies under the tilted base law. All required moments are finite here. Thus $m_k$ increases strictly. An affine function of $m_k$ is nonnegative at every $k$ if and only if it is nonnegative at the two endpoints. Equation~\eqref{eq:deriv} proves the equivalence.
\end{proof}

For the Gaussian credit parametrization, $c(t)=\Phi^{-1}(p(t))$ and \eqref{eq:musigma} give an equivalent pair of inequalities:
\begin{equation}\label{eq:gaussianspeed} \dot c(t)+\frac{\dot\rho(t)}{2(1-\rho(t))}\left[c(t)+\frac{m_k(t)}{\sqrt{\rho(t)}}\right]\geq0,\qquad k\in\{1,n\}. \end{equation}
Thus smooth changes in a marginal default curve place exact restrictions on the speed of changes in dependence. The inequalities characterize horizon curves under the fixed initial information used throughout the paper.

\section{Large portfolios and sharp boundary rates}
\subsection{Compatibility at every portfolio size}
The familiar large-portfolio interpretation of the Gaussian model is that conditional idiosyncratic uncertainty diversifies away; see \citet{vasicek2002}. This gives a general compatibility criterion before any Gaussian calculations are used.

\begin{proposition}\label{prop:infinite}
Let $Q_0,Q_1$ have fixed laws on $[0,1]$, and let $K_{a,n}\mid Q_a\sim\Bin(n,Q_a)$. Then $K_{0,n}\st K_{1,n}$ for every positive integer $n$ if and only if $Q_0\st Q_1$.
\end{proposition}
\begin{proof}
If $Q_0\st Q_1$, couple them by common quantiles so $Q_0\leq Q_1$. Conditional on this pair, use the same independent uniforms to generate the Bernoulli indicators in both counts. This orders the counts for every $n$. Conversely, $K_{a,n}/n$ converges in distribution to $Q_a$, for example by conditional variance and $\E[(K_{a,n}/n-Q_a)^2]\leq1/(4n)$. Stochastic order is preserved under weak limits, as is seen by taking expectations of bounded continuous increasing functions. Hence ordering for every $n$ implies $Q_0\st Q_1$.
\end{proof}

\begin{corollary}\label{cor:rigid}
Let $0<p_0,p_1<1$ and $0<\rho_0,\rho_1<1$. The Gaussian horizon laws are absorbing-compatible at every portfolio size if and only if $\rho_1=\rho_0$ and $p_1\geq p_0$.
\end{corollary}
\begin{proof}
By Proposition~\ref{prop:infinite}, the mixing variables must be stochastically ordered. Their distribution functions have the form in Corollary~\ref{cor:location}, with $F=G=\Phi$. If the scales differ, the affine difference in \eqref{eq:affine} takes both signs because $\Phi^{-1}(q)$ ranges over $\R$. Stochastic ordering is therefore impossible. Equality of scales is equivalent to equality of $\rho$, and with this scale fixed the order is equivalent to $c(p_1)\geq c(p_0)$. Sufficiency also follows directly from the common-factor construction.
\end{proof}

The corollary gives an all-size formulation of the Gaussian rigidity discussed by \citet{andersen2006}. It applies to the prescribed homogeneous Gaussian binomial-mixture horizon laws, with arbitrary temporal couplings allowed for each finite $n$. Within this family, compatibility for all portfolio sizes requires a common latent correlation across horizons.

\subsection{A uniform orthant estimate}
We record the logarithmic estimate needed for the interval boundaries. It is a direct extreme-value argument for an equicorrelated normal probability, a problem treated in greater generality by \citet{rinott2001}.

\begin{lemma}\label{lem:asympt}
Uniformly for $p$ and $\rho$ in compact subsets of $(0,1)$, as $n\to\infty$,
\begin{equation}\label{eq:logA} \log A_n(p,\rho)=-\frac{1-\rho}{\rho}\log n+\frac{c(p)\sqrt{2(1-\rho)}}{\rho}\sqrt{\log n}+O(\log\log n). \end{equation}
\end{lemma}
\begin{proof}
Let $\epsilon_1,\ldots,\epsilon_n,Z$ be independent standard normals, $M_n=\max_i\epsilon_i$, and use $\mu,\sigma$ from \eqref{eq:musigma}. Conditional independence gives
\begin{equation}\label{eq:maxrep} A_n(p,\rho)=\Pp(M_n\leq\mu+\sigma Z)=\E\phibar\!\left(\frac{M_n-\mu}{\sigma}\right). \end{equation}
Put $L=\log n$, $b=\sqrt{2L}$, and $d=3\log L/b$. Standard normal tail bounds give
\begin{equation}\label{eq:maxbounds} \Pp(M_n>b+d)\leq n\phibar(b+d)=O(L^{-7/2}),\qquad \Pp(M_n<b-d)\leq\exp\{-aL^{5/2}\} \end{equation}
for some constant $a>0$ and sufficiently large $n$. For the second inequality use $\Phi(b-d)^n\leq\exp[-n\phibar(b-d)]$ and the lower normal tail bound. These statements do not involve $p$ or $\rho$.

Since $\phibar$ is decreasing, \eqref{eq:maxrep} is bounded below by $\Pp(M_n\leq b+d)\phibar((b+d-\mu)/\sigma)$ and above by $\phibar((b-d-\mu)/\sigma)+\exp\{-aL^{5/2}\}$. On the stated compact sets, $\mu$ is bounded and $\sigma$ is bounded above and away from zero. The additive exponential error is uniformly negligible relative to the normal tail in the upper bound. The normal tail expansion $\log\phibar(x)=-x^2/2-\log x-\tfrac12\log(2\pi)+O(x^{-2})$ therefore yields
\begin{equation}\label{eq:intermediate} \log A_n=-\frac{L}{\sigma^2}+\frac{\mu\sqrt{2L}}{\sigma^2}+O(\log L). \end{equation}
Substitution of \eqref{eq:musigma} proves \eqref{eq:logA}, with a uniform remainder.
\end{proof}

\subsection{The correlation boundary expansion}
\begin{theorem}\label{thm:rate}
Fix $0<p_0<p_1<1$ and $0<\rho_0<\bar\rho<1$. Let $[\ell_n,u_n]$ be the admissible interval of Theorem~\ref{thm:band}, and put
\begin{equation}\label{eq:C} C=\rho_0\sqrt{2(1-\rho_0)}\,[c(p_1)-c(p_0)]>0. \end{equation}
For all sufficiently large $n$, both boundaries are interior to $[0,\bar\rho]$, and
\begin{equation}\label{eq:boundrate} \ell_n=\rho_0-\frac{C}{\sqrt{\log n}}+O\!\left(\frac{\log\log n}{\log n}\right),\qquad u_n=\rho_0+\frac{C}{\sqrt{\log n}}+O\!\left(\frac{\log\log n}{\log n}\right). \end{equation}
In particular, the interval width is $2C/\sqrt{\log n}+O(\log\log n/\log n)$.
\end{theorem}
\begin{proof}
Write $a(r)=(1-r)/r$ and $b(p,r)=c(p)\sqrt{2(1-r)}/r$. For any fixed sufficiently small $\varepsilon>0$, Lemma~\ref{lem:asympt} shows that the sign of
\begin{equation}\label{eq:diff} \log A_n(p_1,r)-\log A_n(p_0,\rho_0)=[a(\rho_0)-a(r)]L+[b(p_1,r)-b(p_0,\rho_0)]\sqrt L+O(\log L) \end{equation}
at $r=\rho_0\pm\varepsilon$ is respectively positive and negative. The same conclusion holds with $(p_0,p_1)$ replaced by $(1-p_0,1-p_1)$. Thus the roots defining both boundaries are interior for large $n$ and converge to $\rho_0$.

For bounded real $h$, substitute $r=\rho_0+h/\sqrt L$ in \eqref{eq:diff}. Since $a'(\rho_0)=-1/\rho_0^2$, Taylor expansion, uniform for bounded $h$, gives
\begin{equation}\label{eq:scaledroot} \log A_n(p_1,\rho_0+h/\sqrt L)-\log A_n(p_0,\rho_0)=\sqrt L\left[\frac{h}{\rho_0^2}+\frac{[c(p_1)-c(p_0)]\sqrt{2(1-\rho_0)}}{\rho_0}\right]+O(\log L). \end{equation}
Taking $h=-C-\eta$ and $h=-C+\eta$ for any fixed $\eta>0$ brackets the lower root for all sufficiently large $n$. Hence $\sqrt L(\ell_n-\rho_0)$ is bounded and converges to $-C$. Substituting this bounded value of $h$ into \eqref{eq:scaledroot} and using equality at the root gives $h=-C+O(\log L/\sqrt L)$, which is the lower expansion.

For the upper boundary use $S_n(p,r)=A_n(1-p,r)$ and $c(1-p)=-c(p)$. The difference of the two $b$ terms changes sign, so the corresponding equation has root $h=C+O(\log L/\sqrt L)$. This proves the upper expansion and the width formula.
\end{proof}

The leading symmetry in \eqref{eq:boundrate} emerges as $n$ grows. At moderate $n$, the upper restriction can be substantially weaker than the lower restriction. The lower boundary is governed by an extreme event whose probability may be very small. Assessing economic materiality therefore also requires the magnitude of the resulting tail-probability changes.

\begin{corollary}\label{cor:critical}
Under the assumptions of Theorem~\ref{thm:rate}, for nonzero $\delta$ sufficiently close to zero let $N_*(\delta)$ be the largest integer portfolio size for which the horizon pairs $(p_0,\rho_0)$ and $(p_1,\rho_0+\delta)$ are absorbing-compatible. Then $N_*(\delta)$ is finite and
\begin{equation}\label{eq:critical} \lim_{\delta\to0,\ \delta\ne0}\delta^2\log N_*(\delta)=C^2. \end{equation}
\end{corollary}
\begin{proof}
Size one is compatible, and Proposition~\ref{prop:nesting} makes the compatible sizes an initial segment. Corollary~\ref{cor:rigid} shows that the segment is finite for $\delta\ne0$. Every fixed size is compatible for all sufficiently small $|\delta|$ by Theorem~\ref{thm:band}; hence $N_*(\delta)\to\infty$.

Fix $0<\eta<C$. For $\delta>0$, the relevant radius is $u_n-\rho_0$, which is asymptotic to $C/\sqrt{\log n}$. For $\delta<0$, use $\rho_0-\ell_n$, with the same asymptotic equivalent. At the integer sizes nearest $\exp[(C-\eta)^2/\delta^2]$ and $\exp[(C+\eta)^2/\delta^2]$, these radii are respectively greater and smaller than $|\delta|$ for sufficiently small $|\delta|$. Nesting places $N_*(\delta)$ between these sizes. Multiply their logarithms by $\delta^2$ and let first $\delta\to0$ and then $\eta\downarrow0$.
\end{proof}

The critical-size law in \eqref{eq:critical} may also be written as $\log N_*(\delta)\sim C^2/\delta^2$. The constant $C$ depends on the earlier latent correlation and on the increase in the normal-transformed cumulative default probability. Consequently, the first portfolio size at which a correlation discrepancy becomes infeasible can change substantially across default-probability curves. The exact interval equations remain relevant for realistic finite pools, where the limiting approximation can be asymmetric.

\subsection{A finite-size certificate}
The all-size result can be made quantitative without a Gaussian assumption. The following bound uses the concentration inequality of \citet{hoeffding1963}.

\begin{proposition}\label{prop:certificate}
Let $K_{a,n}\mid Q_a\sim\Bin(n,Q_a)$. Suppose $z\in(0,1)$ and $0<\varepsilon<\min(z,1-z)$ satisfy $\delta=\Pp(Q_0\geq z+\varepsilon)-\Pp(Q_1\geq z-\varepsilon)>0$. Then
\begin{equation}\label{eq:certificate} \Pp(K_{0,n}/n\geq z)-\Pp(K_{1,n}/n\geq z)\geq\delta-2e^{-2n\varepsilon^2}. \end{equation}
In particular, compatibility fails whenever $n>\log(2/\delta)/(2\varepsilon^2)$.
\end{proposition}
\begin{proof}
Conditionally on $Q_0\geq z+\varepsilon$, the probability that $K_{0,n}/n<z$ is at most $e^{-2n\varepsilon^2}$. Thus $\Pp(K_{0,n}/n\geq z)\geq\Pp(Q_0\geq z+\varepsilon)-e^{-2n\varepsilon^2}$. Splitting on $Q_1\geq z-\varepsilon$ and applying the opposite conditional tail bound gives $\Pp(K_{1,n}/n\geq z)\leq\Pp(Q_1\geq z-\varepsilon)+e^{-2n\varepsilon^2}$. Subtraction proves the claim. The count threshold is equivalently $\lceil nz\rceil$.
\end{proof}

For continuous mixing distributions that fail stochastic order, such a positive $\delta$ exists after choosing a sufficiently small $\varepsilon$ around a strict crossing violation. The bound supplies an explicit sufficient size for incompatibility. In the Gaussian model, the two-event test determines the first failing integer size.

\section{Computational methodology and numerical results}
\subsection{Evaluation of the admissible correlation interval}
The numerical calculations use one-dimensional adaptive quadrature and bracketed scalar root finding. For $n\geq2$, $0<p_0\leq p_1<1$, and $0\leq\rho_0<\bar\rho<1$, define the log probabilities $a_n(p,r)=\log A_n(p,r)$ and $s_n(p,r)=a_n(1-p,r)$. With $\mu=\mu(p,r)$ and $\sigma=\sigma(r)$ from \eqref{eq:musigma}, the orthant integral takes the form
\begin{equation}\label{eq:integration} A_n(p,r)=\frac{1}{\sigma\sqrt{2\pi}}\int_{-\infty}^{\infty}\exp\!\left\{n\log\Phi(x)-\frac{(x-\mu)^2}{2\sigma^2}\right\}\,dx,\qquad r>0.\end{equation}
Writing $h(x)=n\log\Phi(x)-(x-\mu)^2/(2\sigma^2)$ and $x_*=\operatorname*{arg\,max}_{x\in\mathbb R}h(x)$, we evaluate
\begin{equation}\label{eq:shifted} a_n(p,r)=h(x_*)-\log(\sigma\sqrt{2\pi})+\log\!\int_{-\infty}^{\infty}\exp\{h(x)-h(x_*)\}\,dx.\end{equation}
The function $h$ is strictly concave and its mode is the unique root of
\begin{equation}\label{eq:mode} n\frac{\phi(x_*)}{\Phi(x_*)}=\frac{x_*-\mu}{\sigma^2}.\end{equation}
The scaled integrand in \eqref{eq:shifted} is bounded by one. The implementation evaluates the normal log distribution function directly, brackets the mode, and integrates over $[x_*-13\sigma,x_*+13\sigma]$. Strong concavity yields $h(x)-h(x_*)\leq-(x-x_*)^2/(2\sigma^2)$, which bounds the omitted unnormalized integral by $2\sigma\int_{13}^{\infty}\exp(-u^2/2)\,du$. At $r=0$, the analytic value $a_n(p,0)=n\log p$ is used.

The feasible interval is computed through the monotone equations of Theorem~\ref{thm:band}. The following procedure also gives the one-step projection used in the repair examples.

\medskip
\noindent\textbf{Algorithm 1 (Feasible correlation band and projection).}
\begin{enumerate}[leftmargin=2em,label=\arabic*.]
\item Compute $a_0=a_n(p_0,\rho_0)$ and $s_0=s_n(p_0,\rho_0)$ using \eqref{eq:shifted}.
\item Set $\ell_n=0$ if $n\log p_1\geq a_0$; otherwise solve $a_n(p_1,r)=a_0$ on $(0,\rho_0]$ by bracketed root finding.
\item Set $u_n=\bar\rho$ if $s_n(p_1,\bar\rho)\leq s_0$; otherwise solve $s_n(p_1,r)=s_0$ on $[\rho_0,\bar\rho)$.
\item Given a proposed later correlation $\widehat\rho_1$, return $[\ell_n,u_n]$ and $\rho_1^*=\min\{u_n,\max\{\ell_n,\widehat\rho_1\}\}$.
\end{enumerate}
\medskip

The root brackets follow from Lemma~\ref{lem:monotonic}, and each root is unique when its corresponding endpoint inequality is active. Along a horizon grid, repeating Algorithm~1 with the preceding repaired correlation produces the sequence in \eqref{eq:repair}. The integrals and scalar roots determine the feasible interval without numerical integration of the full count distribution. Full-tail integration is reserved for verification and for the diagnostic $V_n$ in \eqref{eq:violation}.

The Python implementation uses absolute and relative quadrature tolerances of $10^{-11}$ and $3\times10^{-11}$ for the mode-centered orthant integral. The scalar roots use a bracketed Brent solver with an absolute tolerance of $5\times10^{-13}$; the mode solver uses $2\times10^{-12}$. The independent full-tail quadrature uses absolute and relative tolerances $10^{-12}$ and $3\times10^{-12}$. For very large mathematical portfolio sizes the computation retains $n$ in the log-integrand formulation, avoiding the underflow that would arise from direct evaluation of small orthant probabilities.

\subsection{Numerical verification}
Two closed-form identities serve as analytic benchmarks. At $p=\rho=1/2$, the mixing variable is uniform on $(0,1)$, so $A_n=1/(n+1)$. At $n=2$ and $p=1/2$, $A_2=1/4+\arcsin(\rho)/(2\pi)$. Nine evaluations of these identities yield a maximum absolute error of $\bencherrorest$ in the probability or log-probability scale, as appropriate. Thirty parameter-pair comparisons at $n=8,25,60$ agree between the two-event criterion and an independent integration of all count-tail probabilities. The program also checks the monotonicity of the repaired survival and all-default probabilities across the five-horizon example.

A secondary accuracy run tightens integration tolerances and expands the mode-centered interval. The largest absolute change in the checked log-orthant probabilities is $\mcapaccuracylog$; the largest absolute change in a checked correlation boundary is $\mcapaccuracyroot$. These are empirical numerical sensitivity diagnostics. All experiments use constructed parameters, and the computations are deterministic.

\subsection{Increasing mean defaults with an incompatible distribution}
Consider a portfolio of $100$ borrowers. The earlier pair is $(p_0,\rho_0)=(0.02,0.10)$ and the later raw pair is $(p_1,\rho_1)=(0.025,0.40)$. Table~\ref{tab:example} reports the two extreme events. Expected defaults rise from $2$ to $2.5$, and the probability of no defaults rises from $0.270717$ to $0.514092$. The probability of at least one default therefore falls by approximately $24.34$ percentage points. This decline violates the stochastic ordering required for an absorbing coupling.

\begin{table}[htbp]
\centering\small
\caption{A constructed incompatibility and its one-step repair, $n=100$. The last column is the base-ten logarithm of the all-default probability.}\label{tab:example}
\begin{tabular}{lrrrrr}
\toprule
Forecast & $p$ & $\rho$ & $\E K$ & $\Pp(K=0)$ & $\log_{10}\Pp(K=n)$\\
\midrule
Earlier & 0.020 & 0.100000 & 2.0 & 0.270717 & -27.430 \\
Later raw & 0.025 & 0.400000 & 2.5 & 0.514092 & -8.146 \\
Later repaired & 0.025 & 0.153059 & 2.5 & 0.270717 & -18.947 \\

\bottomrule
\end{tabular}
\end{table}

\begin{figure}[htbp]
\centering\includegraphics[width=.83\textwidth]{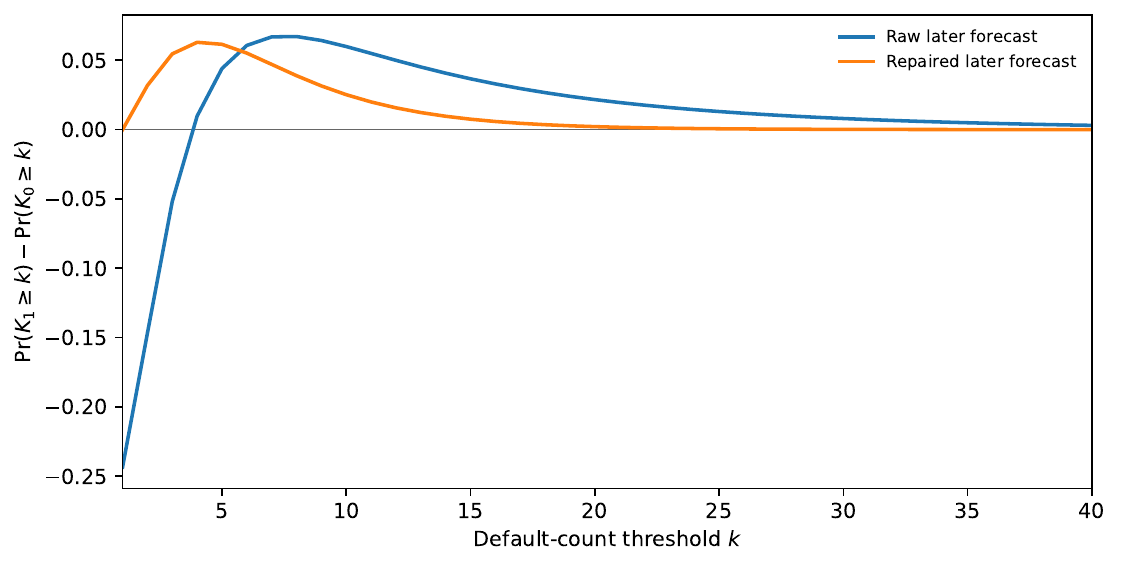}
\caption{Changes in count-tail probabilities in the constructed example. A negative value contradicts absorption. The repaired curve satisfies all $100$ count inequalities; the display focuses on thresholds up to $40$.}\label{fig:tails}
\end{figure}

The closest feasible later correlation is $0.153059$. This keeps $p_1=0.025$ and the expected count $2.5$ unchanged, while making the no-default probability equal to its earlier value. Figure~\ref{fig:tails} shows that every count tail is then nondecreasing. Direct enumeration gives the first incompatibility at $n=\firstfailure$ for this fixed parameter pair.

For scale, consider $100$ equal loans with a deterministic loss of \$1 million per default. The expected loss in the first \$1 million layer equals \$1 million times the probability of at least one default. The raw forecast lowers that expected layer loss by approximately \$243,375 as maturity increases. This illustrates the economic size of the coherence violation in the constructed scenario. The repaired distribution has $V_{100}=0$, with the earlier marginal probability and the later marginal probability maintained at their stated levels.

\subsection{Portfolio size and slow convergence}
For $(p_0,p_1,\rho_0)=(0.02,0.03,0.15)$ and cap $\bar\rho=0.99$, Table~\ref{tab:bands} reports the exact interval boundaries and the scaled distances predicted to converge to $C=\limitconstant$. At $n=100$, the admissible interval is approximately $[0.135019,0.253694]$. At $n=10^6$ it has narrowed to $[0.142354,0.168643]$.

\begin{table}[htbp]
\centering\small
\caption{Exact finite-portfolio intervals and the asymptotic scaling. The two scaled columns converge to $\limitconstant$. Very large sizes are mathematical diagnostics.}\label{tab:bands}
\begin{tabular}{rrrrr}
\toprule
$n$ & $\ell_n$ & $u_n$ & $\sqrt{\log n}(\rho_0-\ell_n)$ & $\sqrt{\log n}(u_n-\rho_0)$\\
\midrule
$10^{1}$ & 0.118262 & 0.495604 & 0.048160 & 0.524429 \\
$10^{2}$ & 0.135019 & 0.253694 & 0.032148 & 0.222524 \\
$10^{3}$ & 0.138823 & 0.198195 & 0.029377 & 0.126669 \\
$10^{6}$ & 0.142354 & 0.168643 & 0.028419 & 0.069293 \\
$10^{12}$ & 0.144503 & 0.159804 & 0.028897 & 0.051536 \\
$10^{40}$ & 0.146837 & 0.154259 & 0.030354 & 0.040875 \\
$10^{160}$ & 0.148344 & 0.151916 & 0.031795 & 0.036777 \\

\bottomrule
\end{tabular}
\end{table}

\begin{figure}[htbp]
\centering\includegraphics[width=\textwidth]{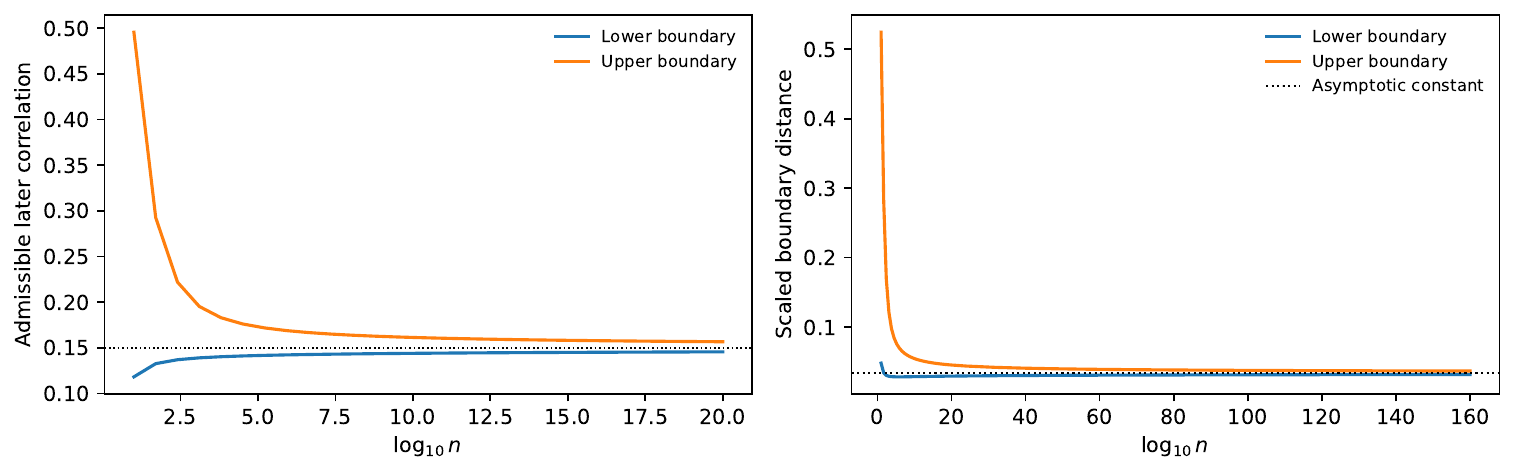}
\caption{Admissible later correlations and their asymptotic scaling for $p_0=0.02$, $p_1=0.03$, $\rho_0=0.15$. The left panel displays sizes through $10^{20}$; the right panel includes much larger mathematical sizes to show the logarithmically slow convergence.}\label{fig:bands}
\end{figure}

The upper and lower scaled distances remain visibly different at $n=10^{12}$. Sizes up to $10^{160}$ serve as mathematical diagnostics of the slow asymptotic convergence. The leading-order formula identifies the contraction rate, while the scalar equations provide exact finite-$n$ boundaries. Figure~\ref{fig:bands} displays the shrinking interval and the scaled endpoints. All reported very-large-size calculations use the log-integrand representation in \eqref{eq:integration} to display convergence of the boundary expansion.

\subsection{Sensitivity to default probabilities and initial correlation}
The shape of the admissible interval depends on the default-probability increment, the earlier correlation, and portfolio size. Table~\ref{tab:sensitivity} reports six parameter configurations, each evaluated at $n=50$ and $n=500$. All scenarios use the same correlation cap $\bar\rho=0.99$ and are constructed independently of observed loan data.

\begin{table}[htbp]
\centering\small
\caption{Sensitivity of the exact admissible correlation interval to cumulative default probabilities, earlier correlation, and portfolio size. The interval width equals $u_n-\ell_n$. All entries are deterministic.}\label{tab:sensitivity}
\begin{tabular}{rrrrrrr}
\toprule
$p_0$ & $p_1$ & $\rho_0$ & $n$ & $\ell_n$ & $u_n$ & Width\\
\midrule
0.010 & 0.015 & 0.10 & 50 & 0.0890 & 0.2734 & 0.1844 \\
0.010 & 0.015 & 0.10 & 500 & 0.0927 & 0.1566 & 0.0638 \\
0.020 & 0.025 & 0.10 & 50 & 0.0929 & 0.1790 & 0.0861 \\
0.020 & 0.025 & 0.10 & 500 & 0.0954 & 0.1263 & 0.0310 \\
0.020 & 0.030 & 0.15 & 50 & 0.1327 & 0.2934 & 0.1608 \\
0.020 & 0.030 & 0.15 & 500 & 0.1380 & 0.2081 & 0.0701 \\
0.050 & 0.065 & 0.15 & 50 & 0.1353 & 0.2263 & 0.0910 \\
0.050 & 0.065 & 0.15 & 500 & 0.1401 & 0.1835 & 0.0434 \\
0.050 & 0.065 & 0.30 & 50 & 0.2762 & 0.3845 & 0.1083 \\
0.050 & 0.065 & 0.30 & 500 & 0.2827 & 0.3458 & 0.0631 \\
0.100 & 0.125 & 0.30 & 50 & 0.2738 & 0.3694 & 0.0956 \\
0.100 & 0.125 & 0.30 & 500 & 0.2813 & 0.3396 & 0.0583 \\

\bottomrule
\end{tabular}
\end{table}

For example, increasing the cumulative probability from $0.020$ to $0.025$ at $\rho_0=0.10$ permits later correlations in approximately $[0.0929,0.1790]$ for $50$ borrowers, and $[0.0954,0.1263]$ for $500$. The upper endpoint moves considerably more than the lower endpoint in this range. The scenarios with higher initial correlations also show material differences in the admissible downward and upward changes. These finite-size asymmetries support direct interval calculation when modelers assess maturity-specific parameter estimates.

\subsection{Several horizons}
Table~\ref{tab:schedule} gives a five-horizon example with increasing individual default probabilities and oscillating raw correlations. The sequential projection keeps each probability unchanged. The repaired no-default probabilities decrease; the all-default probabilities, also checked in the supplied results, increase. Proposition~\ref{prop:repair} therefore gives an absorbing process for the entire repaired grid.

\begin{table}[htbp]
\centering\small
\caption{Sequential repair for a constructed $100$-borrower horizon curve. Raw and repaired survival refer to the probability that no borrower has defaulted.}\label{tab:schedule}
\begin{tabular}{rrrrrr}
\toprule
Horizon & $p$ & Raw $\rho$ & Repaired $\rho$ & Raw survival & Repaired survival\\
\midrule
1 & 0.010 & 0.120 & 0.120000 & 0.498412 & 0.498412 \\
2 & 0.018 & 0.300 & 0.291198 & 0.506242 & 0.498412 \\
3 & 0.026 & 0.080 & 0.269229 & 0.174954 & 0.382275 \\
4 & 0.033 & 0.250 & 0.254583 & 0.300531 & 0.305172 \\
5 & 0.040 & 0.150 & 0.242514 & 0.152422 & 0.244556 \\

\bottomrule
\end{tabular}
\end{table}

The adjustment at the second horizon affects the feasible intervals at all subsequent horizons. The example shows how successive projections deliver a compatible sequence while carrying forward the dependence on earlier choices. Joint calibration across maturities would optimize those choices over the entire grid.

\section{Modeling implications}
The exact interval test can be incorporated into a modular multi-horizon credit-forecasting pipeline. At a common forecast origin, an individual borrower model supplies nondecreasing cumulative default probabilities, and a dependence model supplies an estimated latent correlation at each maturity. The Gaussian count-mixture law follows from these inputs. The two extreme-event probabilities then determine the admissible interval for the next correlation estimate. When a proposed correlation lies outside the interval, a sequential projection produces a coherent sequence while preserving the marginal probabilities.

The test has an interpretation through portfolio loss layers. A count-tail decline at threshold $k$ implies that the expected loss on the corresponding single-default-width layer decreases with maturity. The maximum decline $V_n$ in \eqref{eq:violation} measures the magnitude of this inconsistency across thresholds. In the example in Section~8.3, the most pronounced violation occurs at the first-default threshold and is economically appreciable. Other parameter choices can produce violations concentrated in extremely unlikely events, making the magnitude and location of the breach important alongside the binary feasibility decision.

The construction in Proposition~\ref{prop:coupling} supplies an absorbing process on a prescribed horizon grid. The resulting temporal law is determined by its count quantiles and borrower permutation. Additional dynamics may be imposed according to the intended use, including conditional intensities, transition effects, and forward-starting exposures. The feasible correlation intervals provide restrictions on the horizon marginals that such dynamics must preserve.

Exchangeability and equal default losses make count stochastic order sufficient for the present construction. Heterogeneous borrowers can have the same count distribution while differing in their identity-specific default patterns, and unequal exposures alter the portfolio-loss order. Extensions to these settings require the corresponding multivariate and loss-distribution restrictions. The Gaussian specification also imposes a particular dependence tail structure; the single-crossing result offers a route to other mixing families when their distribution functions satisfy its hypothesis.

An empirical application would follow cumulative first-default outcomes for a fixed initial portfolio across several horizons, with explicit treatment of censoring and competing exits. Horizon-specific cumulative default probabilities and dependence parameters could be estimated from those histories, together with their joint uncertainty. The resulting confidence regions could be checked against the feasible correlation intervals. The deterministic calculations here isolate the structural restrictions and provide reproducible benchmarks for such an implementation.

\section{Conclusion}
A finite homogeneous portfolio admits maturity-dependent Gaussian correlations within an exact interval determined by the probabilities of no defaults and all defaults. Under a single-crossing condition for the mixing distribution functions, these two events characterize the stochastic order of the entire default-count distribution. The intervals are nested in portfolio size and contract towards the earlier correlation with a common leading boundary displacement of order $(\log n)^{-1/2}$. The corresponding critical portfolio size grows exponentially in the inverse squared correlation discrepancy. The coupling, repair, and smooth-curve results provide complementary constructions for maturity-consistent portfolio forecasts. The numerical sensitivity analysis documents the finite-size asymmetry and its variation across default-probability and correlation settings.

\begingroup
\small\raggedright

\endgroup
\end{document}